\UseRawInputEncoding
\documentclass[10pt]{iopart}
\usepackage{graphicx}
\usepackage{amssymb}
\usepackage{iopams} 

\newcommand{\beq}{\begin{equation}}
\newcommand{\eeq}{\end{equation}}
\newcommand{\be}{\begin{equation*}}
\newcommand{\ee}{\end{equation*}}
\newcommand{\beqa}{\begin{eqnarray}}
\newcommand{\eeqa}{\end{eqnarray}}
\newcommand{\bea}{\begin{eqnarray*}}
\newcommand{\eea}{\end{eqnarray*}}

\def\stackunder#1#2{\mathrel{\mathop{#1}\limits_{#2}}}

\newcommand{\bigmean}[1]{\left\langle#1\right\rangle}
\newcommand{\mean}[1]{\langle#1\rangle}

\newcommand{\lap}[1]{\mathrel{\mathop{\cal L}\limits_{#1}^{}}}

\newcommand{\prob}{\mathbb{P}}
\newcommand{\dd}{{\rm d }}

\newcommand{\eq}{\mathrm{eq}}
\newcommand{\taum}{\mean{\tau}}
\newcommand{\ff}{{\varphi}}

\begin{document}

\title{Poisson points, resetting, universality and the role of the last item}

\author{Claude Godr\`eche}

\address{Universit\'e Paris-Saclay, CNRS, CEA, Institut de Physique Th\'eorique,\\
91191 Gif-sur-Yvette, France}

\begin{abstract}
For a stochastic process reset at random times,
we discuss to what extent the probabilities of some orderings of observables associated with the intervals of time between resetting events are universal, i.e., independent of the choice of the observables, and in particular, to what extent universality depends on the choice of the distribution of these intervals.
For Poissonian resetting, universality  relies only on a combinatorial argument and
on the statistical properties of Poisson points.
For a generic distribution of time intervals between resets, universality no longer holds in general.

\end{abstract}

\address{\today}

\eads{\mailto{claude.godreche@ipht.fr}}

\maketitle

\vskip30pt

Consider $n$ iid (independent, identically distributed) continuous random variables $X_1,X_2,\dots,X_n$.
The probability that any of them, say the last one for example, is larger than the $n-1$ remaining ones is equal to $1/n$.
This also holds if these variables are exchangeable, i.e., if a permutation of these variables does not change their joint distribution \cite{kingman}.
Exchangeable random variables are not necessarily independent, while iid random variables are necessarily exchangeable.
Similarly, the probability that these random variables are monotonically increasing, $\prob(X_1<X_2<\cdots<X_n)$, (or decreasing) is equal to $1/n!$.
One can likewise consider events involving orderings of these random variables of increasing complexity, whose probability can be arbitrarily difficult to determine (see, e.g., \cite{jm}).
These probabilities are independent of the common distribution of $X_1,X_2,\dots,X_n$, as long as the latter is continuous.

If now the number of these random variables is no longer fixed but is itself a fluctuating quantity, that henceforth we denote by $N_t$ (with a notation that will shortly appear natural), then the probability of these different events should be averaged with the probability $\prob(N_t=n)$.

In this letter we will expand on the aforementioned topic.
In particular, we will provide an alternative perspective on certain questions raised in the recent work \cite{letter}.
As stated in \cite{letter}, given a random process undergoing stochastic resetting at a constant rate $r$ to a position drawn from some distribution, if the random variables $X_i$ are dynamical observables associated with the intervals between resetting events\footnote{We refer the reader to
\cite{letter} for an introduction to the literature on resetting processes.},
the probabilities of the events oulined above (such as the last element being greater than all preceding elements or the sequence being monotonically increasing) are `super-universal', meaning they are independent of the specific process, the observables $X_i$ and the distribution of the restarting position.

As we now demonstrate, this universality is a simple consequence of the properties of Poisson processes.
Indeed, since the stochastic process is reset at a constant rate $r$\footnote{For simplicity we assume that the process is reset at the origin.},
the resetting events are Poisson points in time, with the following properties \cite{stirz}:
\begin{enumerate}
\item the number of points in $(0,t)$ is given by the Poisson distribution
\beq\label{eq:Poisson}
\prob(N_t=n)=\e^{-rt}\frac{(rt)^n}{n!},
\eeq
\item the $N_t+1$ intervals between these points in $(0,t)$ are statistically equivalent, i.e., exchangeable.
\end{enumerate}

The first property is well known.
The second one is a consequence of a fundamental property of Poisson points, stating that, conditional on $N_t=n$, the positions of these $n$ events in $(0,t)$ are independently and uniformly distributed on $(0,t)$.

Consider now the observables $X_1,X_2,\dots,X_{N_t+1}$
associated with these $N_t+1$ time intervals between resetting events.
For instance, if the stochastic process is Brownian motion, $X$ can be the length $\tau$ of the interval between two resettings, or the maximum attained by the process in this interval, or else the area under the process, and so on.
From the above, it is clear that for each of these cases the corresponding random variables are exchangeable.
Hence, denoting by $Q_1(t)$ the probability that the last observable (be it the interval $\tau$, or the maximum, or the area, etc) is larger than all previous ones, we have
\beq\label{eq:Q1}
\fl
Q_1(t)=\bigmean{\frac{1}{N_t+1}}=\sum_{n\ge0}\frac{\prob(N_t=n)}{n+1}=
\frac{1-\e^{-rt}}{rt}\stackunder{\approx}{t\to\infty}\frac{1}{rt}.
\eeq
Similarly, let $Q_2(t)$ be the probability of the event $\{X_1<X_2<\cdots<X_{N_t+1}\}$.
We have
\beq\label{eq:Q2}
\fl
Q_2(t)=\bigmean{\frac{1}{(N_t+1)!}}=\sum_{n\ge0}\frac{\prob(N_t=n)}{(n+1)!}=
\frac{\e^{-rt}}{\sqrt{rt}}\,I_1(2\sqrt{rt})\stackunder{\approx}{t\to\infty}
\frac{\e^{-rt+2\sqrt{rt}}}{2\sqrt{\pi}(rt)^{3/4}},
\eeq
where $I_n(z)$ is 
the $n$th modified Bessel function of the first kind.
The same applies to the probabilities of more complex orderings of events.

These conclusions were reached in \cite{letter} using a more computational scheme.
The analysis given above is purely combinatorial and relies on the statistical properties of Poisson points.

The natural question to ask at this stage is: what happens if 
the $N_t+1$ intervals in $(0,t)$ are no longer exchangeable?
As is well known, the Poisson process is the simplest renewal process \cite{cox,feller}, where the interarrival times are exponentially distributed with density $\rho(\tau)=r\e^{-r\,\tau}$ and where the number of events $N_t$ in $(0,t)$ is given by the Poisson distribution (\ref{eq:Poisson}).
For any renewal process, the intervals of times between events obey the sum rule
\beq\label{eq:sumr}
\tau_1+\tau_2+\cdots+\tau_{N_t}+B_t=t,
\eeq
where, $B_t$, the last interval between the observation time $t$ and the last event to its left, named the backward recurrence time, is, in the particular case of a Poisson process, statistically equivalent to the $N_t$ intervals $\tau_1,\dots,\tau_{N_t}$.
This property was used above to demonstrate the universality of the results (\ref{eq:Q1}), (\ref{eq:Q2}), regardless of the choice of the observable $X$.
For a generic renewal process however, i.e., for a generic density $\rho(\tau)$ of the intervals of time between resetting events, this property no longer holds, namely the last interval is no longer statistically independent of the $N_t$ previous ones.
One is therefore naturally led to investigate what remains of the universality property mentioned above.

As a test bed, let us focus on the probability that the last observable is the largest one, applied, firstly, to the case where the observable in question is the interval length itself, 
\beq\label{eq:Q1tau}
Q_1^{(\tau)}(t)=\prob\big(B_t>\max(\tau_1,\dots,\tau_{N_t})\big),
\eeq
where the superscript in the notation of this probability enhances the fact that the latter depends on the choice of the observable, as we now show.
The analysis of this case was given in \cite {max-renew}, with the following result, in Laplace space,
\beq\label{eq:laptQ1tau}
\lap{t}Q_1^{(\tau)}(t)=\hat Q_1^{(\tau)}(s)
=\int_0^\infty\dd b\, \frac{\e^{-s b}\int_b^\infty\dd\tau\,\rho(\tau)}{1-\int_0^{b}\dd \tau\rho(\tau)\e^{-s\tau}}.
\eeq
If $\rho(\tau)=r\,\e^{-r\,\tau}$, the result (\ref{eq:Q1}) is recovered \cite{max-renew}.
If $\rho(\tau)$ is uniform $\mathcal{U}(0,1)$, it is found that (see \cite{max-renew})
\beq\label{eq:Q1unif}
Q_1^{(\tau)}(t)\stackunder{\approx}{t\to\infty}\frac{1}{2\,t^2},
\eeq
which is different from the naive estimate $Q_1^{(\tau)}(t)\approx \mean{1/(N_t+1)}$, decaying as $1/2t$ when $t\to\infty$.

If observables $X_1,X_2,\dots$ are attached to the intervals $\tau_1,\tau_2,\dots$, skipping all details, (\ref{eq:laptQ1tau}) is generalised into
\beq\label{eq:resQIx}
\lap{t}Q_1^{(X)}(t)=\int_0^\infty\dd y\,
\frac{\hat\psi(y,s)}{1-\int_{0}^{y}\dd x\,\hat \ff(x,s)},
\eeq
with the definitions
\be
\hat \ff(x,s)={\int_0^\infty\dd\tau\,\e^{-s\tau}\rho(\tau)f(x|\tau)},
\quad 
\hat\psi(y,s)=\int_0^\infty\dd{b}\,\e^{-s{b}}p({b})f(y|{b}),
\ee
where $f(x|\tau)=\frac{\dd}{\dd x}\prob(X<x|\tau)$ is the conditional density of $X$ given $\tau$, and $p(b)=\int_b^\infty\dd\tau\,\rho(\tau)$.
One can check on these expressions that if $X\equiv\tau$, then (\ref{eq:resQIx}) reduces to (\ref{eq:laptQ1tau}).
On the other hand, analysing (\ref{eq:resQIx}) in all generality is not a simple task, therefore we shall resort to a heuristic argument in order to obtain an asymptotic estimate of $Q_1^{(X)}(t)$.

Consider the $N_t$ observables $X_1,X_2\dots$, obtained by taking some random function of the intervals $\tau_1,\dots,\tau_{N_t}$.
For example,
\be
X_1=\tau_1^a\,\zeta_1,\dots,\;
X_{N_t}=\tau_{N_t}^a\,\zeta_{N_t},
\ee
where $a$ is a positive exponent and $\zeta_1,\zeta_2,\dots$ are iid random variables, independent of $\tau_1,\tau_2,\dots$, whose common density $\rho(\tau)$ is taken uniform 
$\mathcal{U}(0,1)$, as in the example leading to (\ref{eq:Q1unif}).
Likewise, we associate the random variable $B_t^a\,\zeta_{N_t+1}$ to the last interval $B_t$.
Furthermore, we define $X_{N_t+1}=\tau_{N_t+1}^{a}\,\zeta_{N_t+1}$, where the interval $\tau_{N_t+1}$ straddles the observation time $t$, and the excess time $E_t$ by
$B_t+E_t=\tau_{N_t+1}$ \cite{cox,feller}.
The questioÒn is whether the probability
\beq\label{eq:Q1unif-def}
Q_1^{(X)}(t)=\prob(B_t^a\,\zeta_{N_t+1}>\max(X_1,\dots,X_{N_t}))
\eeq
is the same as
(\ref{eq:Q1tau}) or at least has the same asymptotic behaviour (\ref{eq:Q1unif}).
A heuristic argument, based on extreme value statistics, and confirmed by numerical simulations, shows that, if the common distribution of $\zeta_1,\zeta_2,\dots$ is exponential, or Gaussian, or more generally is in the Gumbel class \cite{gumbel}, then
\beq\label{eq:Q1unifA}
Q_1^{(X)}(t)\stackunder{\approx}{t\to\infty}\frac{c}{a\,t\,\ln t},
\eeq
which is quite different from the result (\ref{eq:Q1unif}).
The constant $c$ appearing in (\ref{eq:Q1unifA})
depends on the choice of distribution for the random variables $\zeta$.

The heuristic argument is as follows. 
Let for example $\zeta_1,\zeta_2,\dots$ be exponentially distributed random variables with unit parameter.
Let us fix $N_t=n$, and
let us denote by $X_{(1)}$ the largest amongst the $n+1$ random variables $X_1,\dots,X_{n+1}$, and by $X_{(2)}$ the second largest one. 
When the event $\{B_t^a\,\zeta_{n+1}>\max(X_1,\dots,X_n)\}$ holds, then necessarily
$X_{(1)}=X_{n+1}\equiv \tau^a_{n+1} \zeta_{n+1}$, which occurs with probability $1/(n+1)$, and $X_{(2)}=\max(X_1,\dots,X_n)$.
Given this holds, we have the following equivalent inequalities
\beqa\label{eq:equiv}
\fl
B_t^a\,\zeta_{n+1}>X_{(2)}\Longleftrightarrow (\tau_{n+1}-E_t)^a \zeta_{n+1}>X_{(2)}
\Longleftrightarrow \Big(1-\frac{E_t}{\tau_{n+1}}\Big)^a X_{(1)}>X_{(2)}.
\eeqa
We note that in order for $X_{(1)}$ and $X_{(2)}$ to be the largest observables, asymptotically, i.e., when $n$ and $t$ are large, the intervals attached to them are necessarily close to unity, which entails that $E_t\ll 1$, as well as $X_{(1)}\approx \zeta_{(1)}$, $X_{(2)}\approx \zeta_{(2)}$.
Therefore the rightmost inequality in (\ref{eq:equiv}) can be rewritten as
\be
1-a\frac{E_t}{\tau_{n+1}}>\frac{X_{(2)}}{ X_{(1)}}\approx \frac{\zeta_{(2)}}{\zeta_{(1)}},
\ee
which leads finally, in the asymptotic regime, to the equivalence
\beq\label{eq:Einf}
B_t^a\,\zeta_{n+1}>X_{(2)}\Longleftrightarrow E_t<\frac{\zeta_{(1)}-\zeta_{(2)}}{a\, \zeta_{(1)}}.
\eeq
At large times the distribution of $E_t$ becomes stationary, i.e., $E_t\to E_{\eq}$, with (see e.g., \cite{gl})
\beq\label{eq:Eeq}
\prob(E_{\eq}<x)=\frac{1}{\taum}\int_0^x\dd e\,(1-e)
\approx \frac{x}{\taum},
\eeq
because $x$ (playing the role of the right-hand side of (\ref{eq:Einf})) is small (see (\ref{eq:estimGG+})).
So
\beq\label{eq:Bta}
\prob(B_t^a\, \zeta_{n+1}>X_{(2)})\approx \frac{\zeta_{(1)}-\zeta_{(2)}}{a\taum\zeta_{(1)}}.
\eeq
Using a well-known extreme value statistics argument, we have, for large $n$,
\be
\e^{-\zeta_{(1)}}\approx \frac{\xi_1}{n},\qquad \e^{-\zeta_{(2)}}\approx \frac{\xi_1+\xi_2}{n},
\ee
where $\xi_1$ and $\xi_2$ are exponentially distributed with parameter unity, hence
\beq\label{eq:estimGG+}
\frac{\zeta_{(1)}-\zeta_{(2)}}{\zeta_{(1)}}\approx \frac{\ln\frac{\xi_1+\xi_2}{\xi_1} }{\ln n}.
\eeq
Averaging this expression on $\xi_1$ and $\xi_2$, we finally get, after multiplication of (\ref{eq:Bta}) by $1/(n+1)$ and replacement of $n$ by $t/\taum$,
\be
Q_1^{(X)}(t)
\stackunder{\approx}{t\to\infty}\frac{1}{a\,t\ln t},
\ee
which is (\ref{eq:Q1unifA}) with $c=1$.
This very reasoning can be extended to other distributions of the random variables $\zeta_1,\zeta_2,\dots$.
For instance if the latter variables are uniform $\mathcal{U}(0,1)$, one obtains
\beq\label{eq:Qunifunif}
Q_1^{(X)}(t)
\stackunder{\approx}{t\to\infty}\sqrt{\frac{\pi}{16\,t^3}},
\eeq
which again is different from (\ref{eq:Q1unif}).

Comparing (\ref{eq:Q1unif}) to (\ref{eq:Q1unifA}) and (\ref{eq:Qunifunif}) shows that the degeneracy (independence in the choice of observable) holding for Poissonian resetting (see (\ref{eq:Q1})) is lifted, i.e., the probability $Q_1^{(X)}(t)$ now depends on the choice of the observable $X$, with different expressions according to whether $X\equiv \tau$ or $X\equiv \tau^{a}\zeta$, as soon as $\rho(\tau)$ is no longer exponential.

The question of 
whether universality extends beyond the Poissonian resetting protocol, i.e., with a non-exponential 
distribution $\rho(\tau)$
was raised in \cite{letter}, where it was anticipated that this universality would still hold, however only for large $t$, and provided $\rho(\tau)$ decays sufficiently fast at large $\tau$. 
As demonstrated above, this does not hold true in general.
A complete discussion of universality beyond this point would require a thorough investigation.
At the very least, the analysis given above demonstrates the role played by the last item (interval or observable attached to the latter) for the determination of $Q_1^{(X)}(t)$.
For probabilities of other orderings of $X_1,X_2,\dots$ (such as for example $Q_2^{(X)}(t)$), the role of the last element is probably all the less important 
as the definitions of the corresponding events give less weight to this last element.

As a corollary of the above, a way of restoring exchangeability at any finite time consists in considering the $N_t$ first intervals only, discarding the last one, $B_t$.
Now, since these intervals are statistically equivalent, the same holds for any observable associated with these intervals.
Hence the universality property holding with Poisson points for the complete sequence $X_1,\dots,X_{N_t+1}$ now holds for the restricted sequence $X_1,\dots,X_{N_t}$
\textit{regardless of the choice of density $\rho(\tau)$}.
This is illustrated below on a few examples.

Consider first the probability that the last observable of the sequence is the largest one, denoted by $q_1(t)$\footnote{This probability was denoted by $Q^{\mathrm{III}}(t)$ in \cite{max-renew}.}.
Then, for any distribution of the intervals $\rho(\tau)$, and for any choice of the observables attached to these intervals, we have
\be
q_1(t)=\bigmean{\frac{1}{N_t}},
\ee
(where $N_t>0$).
For example, for Poisson points, with $\rho(\tau)=\e^{-r\tau}$, this probability was computed in \cite{max-renew}, with the result
\be
q_1(t)=\e^{-rt}\int_0^{rt}\dd u\,\frac{\e^{u}-1}{u}
\stackunder{\approx}{t\to\infty}\frac{1}{rt},
\ee
which has the same asymptotic behaviour as in (\ref{eq:Q1}).
This asymptotic behaviour also holds for any distribution of the intervals with a finite first moment \cite{max-renew}.
For a generic density $\rho(\tau)$, in Laplace space, we have \cite{max-renew},
\be
\lap{t}q_1(t)=-\frac{1-\hat\rho(s)}{s}\ln(1-\hat\rho(s)).
\ee
As an application, let us consider a simple example of a distribution $\rho(\tau)$ with a fat tail, namely such that $\hat \rho(s)=\e^{-\sqrt{s}}$.
We obtain
\be
q_1(t)\stackunder{\approx}{t\to\infty}\frac{\ln(4t)+\gamma}{2\sqrt{\pi t}},
\ee
(see \cite{max-renew}) where $\gamma$ is Euler's constant.

Likewise, the probability of the event $\{X_1<X_2<\cdots<X_{N_t}\}$ is, for any distribution of the intervals $\rho(\tau)$,
\be
q_2(t)=\bigmean{\frac{1}{N_t!}},
\ee
hence, for Poisson points, for example,
\be
q_2(t)=\e^{-rt}I_0(2\sqrt{rt})
\stackunder{\approx}{t\to\infty}
\frac{\e^{-rt+2\sqrt{rt}}}{2\sqrt{\pi}(rt)^{1/4}},
\ee
which decays slightly slower than (\ref{eq:Q2}).
For a generic density $\rho(\tau)$, in Laplace space, we have 
\be
\lap{t}q_2(t)=-\frac{1-\hat\rho(s)}{s}\,\e^{\hat\rho(s)}.
\ee
For the example of $\hat \rho(s)=\e^{-\sqrt{s}}$,
we obtain
\be
q_2(t)\stackunder{\approx}{t\to\infty}\frac{\e}{\sqrt{\pi t}}.
\ee
In Laplace space, more complicated expressions would be obtained for the case of more complex orderings of events.

Other considerations on the role of the last interval in renewal processes can be found in \cite{max-renew}.
As a final comment, the role of the last item, highlighted in the present letter, is reminiscent of the role of boundary conditions in the language of statistical physics.

\end{document}